\documentclass[numberedappendix]{emulateapj} 
\newcommand{\bl}[1]{\mbox{\boldmath$ #1 $}}

\slugcomment{Accepted for publication by The Astrophysical Journal}
\shorttitle{Properties of embedded  disks}
\shortauthors{Vorobyov} 

\begin{document}

\title{Embedded protostellar disks around (sub-)solar stars. II. Disk masses, sizes,
densities, temperatures and the planet formation perspective}
\author{Eduard I. Vorobyov\altaffilmark{1,}\altaffilmark{2}}
\altaffiltext{1}{Institute for Computational Astrophysics, Saint Mary's University,
Halifax, B3H 3C3, Canada; vorobyov@ap.smu.ca.} 
\altaffiltext{2}{Research Institute of Physics, Southern Federal University, Stachki 194, Rostov-on-Don, 
344090, Russia.} 


\begin{abstract}
We present basic properties of protostellar disks in the embedded phase of star formation (EPSF), 
which is difficult to probe observationally using available observational facilities. We use numerical
hydrodynamics simulations of cloud core collapse and focus on disks formed around stars
in the $0.03-1.0~M_\sun$ mass range. Our obtained disk masses scale near-linearly with 
the stellar mass. The mean and median
disk masses in the Class~0 and I phases ($M_{\rm d,C0}^{\rm mean}=0.12~M_\odot$, 
$M_{\rm d,C0}^{\rm mdn}=0.09~M_\odot$ and $M_{\rm d,CI}^{\rm mean}=0.18~M_\odot$, 
$M_{\rm d,CI}^{\rm mdn}=0.15~M_\odot$, respectively) are greater than those inferred from 
observations by (at least) a factor of 2--3.
We demonstrate that this disagreement may (in part) be caused by the optically thick 
inner regions of protostellar disks, which do not contribute to millimeter dust flux. 
We find that disk masses and surface densities start to systematically exceed that of the minimum
mass solar nebular for objects with stellar mass as low as $M_\ast=0.05-0.1~M_\odot$.
Concurrently, disk radii start to grow beyond 100~AU, 
making gravitational fragmentation in the disk outer regions possible.
Large disk masses, surface densities, and sizes suggest that giant planets may start forming as 
early as in the EPSF, either by means of 
core accretion (inner disk regions) or direct gravitational instability (outer disk regions), thus 
breaking a longstanding stereotype that the planet formation process begins in the Class~II phase. 
\end{abstract}


\section{Introduction}
Theoretical and numerical studies indicate that 
specific physical conditions are needed for planets to form in circumstellar disks
and understanding disk properties may therefore help to choose between competing planet formation
scenarios. In the past decade, much effort has been made to survey circumstellar disks 
around low-mass stars and brown dwarfs in the nearby star-forming regions 
\citep[e.g.][and many others]{Looney03,Vicente05,Andrews05,
Scholz06, Andrews07,Eisner08, Andrews09,Mann09,Furlan09}.
However, these studies have mostly focused on the late evolution stage when the disk, if favorably
located and inclined, is  accessible for observations. On the contrary, the early embedded stage, 
when the disk is shrouded
by a natal cloud core, is difficult to probe with the modern observational techniques and only a handful
of attempted studies have been done so far \citep[e.g.][]{Eisner05,Chiang08,Jorgensen09}. 
In this context, a numerical investigation into the properties of protostellar disks in the 
embedded phase of star formation (hereafter, EPSF) is of considerable interest. 

While there are plenty of theoretical and numerical studies addressing stability, fragmentation,
and mass transport in circumstellar disks \citep[e.g.][]{Laughlin94,Boss98,Gammie01,Johnson03,
Lodato04,Boley06,Durisen07,Mayer07,Stamatellos09,Rafikov09,Clarke09,Rice09}, 
other disk properties such as masses, sizes, typical temperatures and densities have 
received considerably 
less attention owing to the (technical) difficulty with forming circumstellar disks 
self-consistently in numerical hydrodynamics simulations. 
Indeed, numerical modeling of {\it isolated} disks cannot provide
reliable information on, for example, disk masses and sizes because they
depend on the initial masses and rotation rates of parent cloud cores. 

On the other hand, numerical simulations that do form disks self-consistently often focus on 
the radial gas surface density and temperature profiles \citep{Lin90,Hueso05,Rice10,Vor10b},
accretion properties  \citep[e.g.][]{Dullemond06,VB09}, and time evolution of disk and 
stellar masses \citep{Nakamoto94}. In addition, they often adopt many simplifying assumptions 
such as the barotropic equation of state, disk axisymmetry, etc. Three-dimensional numerical 
hydrodynamics simulations often suffer from a small number of considered models due to an enormous
computational load \citep[e.g.][]{Machida10,Kratter10}.

In this paper, we perform a comprehensive numerical analysis of embedded disks, which, 
for the first
time, includes obtaining the typical disk masses, radii, temperatures, and densities 
in the Class 0 and I phases of star formation for a wide spectrum of initial cloud core masses
and angular momenta. We utilize two-dimensional numerical hydrodynamics simulations
in the thin-disk approximation with an accurate treatment of disk thermodynamics. This allows us
to model the formation and long-term evolution of protostellar disks and derive useful
statistical relations between various disk properties and stellar masses and also make suggestions
about the giant planet formation perspective in the EPSF.
This is the third paper in a series focusing on the properties of circumstellar disks 
around (sub-)solar mass stars in the embedded phase of star formation. 
Two previous papers were mainly dedicated to the stability, fragmentation, 
accretion properties \citep{VB10b}, and also to the radial structure of protostellar disks \citep{Vor10b}.

The paper is organized as follows. In Section~\ref{model} we provide a brief description of our numerical
model. Section~\ref{initial} summarizes the initial parameters of cloud cores.
Section~\ref{scheme} provides details on the adopted classification scheme for young stellar objects
and also on the procedure for distinguishing between the disk and infalling envelope.
The main results are presented in Section~\ref{diskproperties} and are summarized in Section~\ref{summary}.
The model caveats are discussed in Section~\ref{caveats}.

\section{Model description}
\label{model}
The main concepts of our approach are explained in detail in \citet{VB10b}
and are briefly reviewed below. We make use of numerical hydrodynamics simulations in 
the thin-disk approximation to compute 
the gravitational collapse of rotating, gravitationally unstable cloud cores. This approximation 
is an excellent means to calculate the evolution for many orbital periods and many model 
parameters. We start our numerical integration in the pre-stellar phase, which is 
characterized by a collapsing {\it starless} cloud core, 
continue into the embedded phase of star formation, during which
a star, disk, and envelope are formed, and terminate our simulations in the T Tauri phase,
when most of the envelope has accreted onto the forming star/disk system.
In the EPSF, the disk occupies the innermost region of our numerical grid, while the 
larger outer part of the grid is taken up by the infalling envelope, 
the latter being the remnant of the parent cloud core. 
This ensures that the protostellar disk is not isolated in the EPSF but is exposed to intense
mass loading from the envelope. In addition, the mass accretion rate onto 
the disk $\dot{M}_{\rm env}$ is not a free parameter of the model 
but is self-consistently determined by the gas dynamics in the envelope.

We introduce a ``sink cell'' at $r_{\rm sc}=6$~AU and impose a free outflow condition on both 
boundaries so that the matter is allowed to flow out of the computational domain but 
is prevented from flowing in. We monitor the gas density in the sink cell and 
when its value exceeds a critical value for the transition from 
isothermal to adiabatic evolution ($\sim 10^{11}$~cm$^{-3}$), we introduce a central point-mass star.
In the subsequent evolution, 90\% of the gas that crosses the inner boundary 
is assumed to land onto the central star plus the inner axisymmetric disk at $r<6$~AU. 
The other 10\% of the accreted gas is assumed to be carried away with protostellar jets. 

Main physical processes that are taken into account in our modeling include 
stellar irradiation, background irradiation with temperature $T_{\rm bg}=10$~K, 
viscous and shock heating, radiative cooling from the
disk surface, and also disk self-gravity. 
The corresponding equations of mass, momentum, and energy transport are

\begin{equation}
\label{cont}
\hskip -5 cm \frac{{\partial \Sigma }}{{\partial t}} =  - \nabla_p  \cdot 
\left( \Sigma \bl{v}_p \right),  
\end{equation}
\begin{eqnarray}
\label{mom}
\frac{\partial}{\partial t} \left( \Sigma \bl{v}_p \right) &+& \left[ \nabla \cdot \left( \Sigma \bl{v_p}
\otimes \bl{v}_p \right) \right]_p =   - \nabla_p {\cal P}  + \Sigma \, \bl{g}_p + \\ \nonumber
& + & (\nabla \cdot \mathbf{\Pi})_p, 
\label{energ}
\end{eqnarray}
\begin{equation}
\frac{\partial e}{\partial t} +\nabla_p \cdot \left( e \bl{v}_p \right) = -{\cal P} 
(\nabla_p \cdot \bl{v}_{p}) -\Lambda +\Gamma + 
\left(\nabla \bl{v}\right)_{pp^\prime}:\Pi_{pp^\prime}, 
\end{equation}
where subscripts $p$ and $p^\prime$ refer to the planar components $(r,\phi)$ 
in polar coordinates, $\Sigma$ is the mass surface density, $e$ is the internal energy per 
surface area, 
${\cal P}=\int^{h}_{-h} P dh$ is the vertically integrated
form of the gas pressure $P$, $h$ is the radially and azimuthally varying vertical scale height
determined in each computational cell using an assumption of local hydrostatic equilibrium,
$\bl{v}_{p}=v_r \hat{\bl r}+ v_\phi \hat{\bl \phi}$ is the velocity in the
disk plane, $\bl{g}_{p}=g_r \hat{\bl r} +g_\phi \hat{\bl \phi}$ is the gravitational acceleration 
in the disk plane, and $\nabla_p=\hat{\bl r} \partial / \partial r + \hat{\bl \phi} r^{-1} 
\partial / \partial \phi $ is the gradient along the planar coordinates of the disk. 

Viscosity enters the basic equations via the viscous stress tensor $\mathbf{\Pi}$. 
We parameterize the magnitude of kinematic viscosity $\nu$ using a modified form 
of the $\alpha$-prescription 
\begin{equation}
\nu=\alpha \, c_{\rm s} \, h \, {\cal F}_{\alpha}(r), 
\end{equation}
where $c_{\rm s}^2=\gamma {\cal P}/\Sigma$ is the square of effective sound speed
calculated at each time step from the model's known ${\cal P}$ and $\Sigma$. The 
function ${\cal F}_{\alpha}(r)=
2 \pi^{-1} \tan^{-1}\left[(r_{\rm d}/r)^{10}\right]$ is a modification to the usual 
$\alpha$-prescription that guarantees that the turbulent viscosity operates 
only in the disk and quickly reduces to zero beyond the disk radius $r_{\rm d}$.
In this paper, we use a spatially and temporally 
uniform $\alpha$, with its value set to $5\times 10^{-3}$.

Radiative cooling from the disk surface is determined using the diffusion
approximation of the vertical radiation transport in a one-zone model of the vertical disk 
structure \citep{Johnson03}
\begin{equation}
\Lambda={\cal F}_{\rm c}\sigma\, T^4 \frac{\tau}{1+\tau^2},
\end{equation}
where $\sigma$ is the Stefan-Boltzmann constant, $T$ is the midplane temperature of gas, 
and ${\cal F}_{\rm c}=2+20\tan^{-1}(\tau)/(3\pi)$ is a function that 
secures a correct transition between the cooling function (from both surfaces of the disk) 
in the optically thick regime $\Lambda_{\rm thick}=16\, \sigma \, T^4/3\tau$ 
and the optically thin one $\Lambda_{\rm thin}=2\,\sigma \,T^4\,\tau$.  We use 
frequency-integrated opacities of \citet{Bell94}.

The heating function is expressed as
\begin{equation}
\Gamma={\cal F}_{\rm c}\sigma\, T_{\rm irr}^4 \frac{\tau}{1+\tau^2},
\end{equation}
where $T_{\rm irr}$ is the irradiation temperature at the disk surface 
determined by the stellar and background black-body irradiation as
\begin{equation}
T_{\rm irr}^4=T_{\rm bg}^4+\frac{F_{\rm irr}(r)}{\sigma},
\label{fluxCS}
\end{equation}
where $T_{\rm bg}$ is the uniform background temperature (in our model set to the 
initial temperature of the natal cloud core)
and $F_{\rm irr}(r)$ is the radiation flux (energy per unit time per unit surface area) 
absorbed by the disk surface at radial distance 
$r$ from the central star. The latter quantity is calculated as 
\begin{equation}
F_{\rm irr}(r)= A_{\rm irr}\frac{L_\ast}{4\pi r^2} \cos{\gamma_{\rm irr}},
\end{equation}
where $\gamma_{\rm irr}$ is the incidence angle of 
radiation arriving at the disk surface at radial distance $r$, $A_{\rm irr}$ is a time-dependent
factor that is introduced to account for the possible attenuation of stellar radiation by the 
envelope\footnote{The effect of this attenuation factor on the disk 
evolution is found to be insignificant, since $A_{\rm irr}$ quickly approaches unity with time.
For a typical run, $A_{\rm irr}\approx0.75$ at $t=0.1$~Myr 
after the formation of the central star and $A_{\rm irr}\approx0.95$ at $t=0.3$~Myr. }, and 
$L_\ast$ is the sum of the accretion luminosity $L_{\rm \ast,accr}$ arising from the gravitational 
energy of accreted gas and
the photospheric luminosity $L_{\rm \ast,ph}$ due to gravitational compression and deuterium burning
in the star interior.

Viscous heating operates in the disk interior and is calculated using the standard expression 
$(\nabla \bl{v})_{pp^\prime} : \mathbf{\Pi}_{pp^\prime}$. 
Heating due to shock waves is taken into account via compressional heating 
${\cal P}\left(\nabla_p \cdot \bl{v}_p \right)$ and 
artificial viscosity.
The vertically integrated gas pressure ${\cal P}$ and internal energy per surface area $e$ are 
related via the ideal 
equation of state ${\cal P}=(\gamma-1)\, e$, with the ratio of specific heats $\gamma=7/5$.

Equations~(\ref{cont})--(\ref{energ}) 
are solved using the method of finite differences with a time-explicit, operator-split 
solution procedure in polar coordinates $(r, \phi)$ on a numerical grid with
$512 \times 512$ grid zones. Advection is treated using the van Leer interpolation scheme.
The update of the internal energy per surface area 
$e$ due to cooling $\Lambda$ and heating $\Gamma$
is done implicitly using the Newton-Raphson method of root finding, complemented by the bisection method
where the Newton-Raphson iterations  fail to converge. 
The radial points are logarithmically spaced.
The innermost grid point is located at the position of the sink cell $r_{\rm sc}=6$~AU, and the 
size of the first adjacent cell varies in the 0.07--0.1~AU range depending on the cloud core 
size.  This corresponds to the radial resolution of $\triangle r$=1.1--1.6~AU at 100~AU. 
More details can be found in \citet{VB10b}.



\section{Initial conditions}
\label{initial}
Initially isothermal ($T_{\rm init}\equiv T_{\rm bg}=10$~K) cloud cores have surface densities 
$\Sigma$ and angular velocities $\Omega$ typical for a collapsing, axisymmetric, magnetically
supercritical core \citep{Basu97}
\begin{equation}
\Sigma={r_0 \Sigma_0 \over \sqrt{r^2+r_0^2}}\:,
\label{dens}
\end{equation}
\begin{equation}
\Omega=2\Omega_0 \left( {r_0\over r}\right)^2 \left[\sqrt{1+\left({r\over r_0}\right)^2
} -1\right],
\label{omega}
\end{equation}
where $\Omega_0$ is the central angular velocity and 
$r_0$ is the radius of central near-constant-density plateau defined 
as $r_0 = \sqrt{A} c_{\rm s}^2 /(\pi G\Sigma_0)$. 
With this choice of $r_0$,  equation~(\ref{dens}) at large radii $r\gg r_0$ leads to 
the gas volume density distribution
$\rho =A c_{\rm s}^2/(2 \pi G r^2)$, if the latter is
integrated in the vertical direction assuming a local vertical hydrostatic equilibrium, 
i.e., $\rho = \Sigma/(2h)$ and $h=c_{\rm s}^2/(\pi G \Sigma)$. This means that our initial 
gas surface density configuration can be considered to have a factor of $A$ 
positive density enhancement compared to
that of the singular isothermal sphere $\rho_{\rm SIS} =c_{\rm s}^2/(2\pi G r^2)$.
Throughout the paper, we use $A=1.2$.


\begin{table}
\caption{Model parameters}
\label{table1}
\begin{tabular}{cccccc}
\hline\hline
Model set & $\beta$ & $\Omega_0$ & $r_0$ & $M_{\rm cl}$  & N \\
\hline
 1 & $2.7\times 10^{-3}$ & 0.5-2.0 & 960-3940  & 0.4-1.8 &  6 \\
 2 & $5.6\times 10^{-2}$ & 0.7-6.0 & 445-3770  & 0.2-1.7 &  8 \\
 3 & $1.3\times 10^{-2}$ & 1.2-12  & 340-3430  & 0.15-1.5 &  8 \\
 4 & $2.3\times 10^{-2}$ & 2.1-29 & 190-2570  & 0.085-1.2 &  7  \\
 \hline
\end{tabular} 
\tablecomments{All masses are in $M_\sun$, distances in AU, and angular velocities
in km~s$^{-1}$~pc$^{-1}$.}
\end{table} 

We present results from 4 model sets, the parameters of which are summarized
in Table~\ref{table1}. Every model set has a distinct 
ratio $\beta=E_{\rm rot}/|E_{\rm grav}|$ of the rotational to gravitational energy calculated as
\begin{equation}
E_{\rm rot}= 2 \pi \int \limits_{r_{\rm sc}}^{r_{\rm
out}} r a_{\rm c} \Sigma \, r \, dr, \,\,\,\,\,\
E_{\rm grav}= - 2\pi \int \limits_{r_{\rm sc}}^{\rm r_{\rm out}} r
g_r \Sigma \, r \, dr.
\label{rotgraven}
\end{equation}
Here, $a_{\rm c} = \Omega^2 r$ and $g_{\rm r}$ are the centrifugal and gravitational accelerations,
respectively, and $r_{\rm out}$
is the core's outer radius. The adopted values of $\beta$ lie within
the limits inferred by \citet{Caselli} for dense molecular cloud cores, $10^{-4}$--$7\times 10^{-2}$.
In addition, every model set is characterized by a distinct ratio $r_{\rm out}/r_0=6$ 
in order to generate gravitationally unstable truncated cores of similar form. 
As a result, individual cloud cores within each set of models have
equal $\beta$ and $r_{\rm out}/r_0$ but distinct
masses $M_{\rm cl}$, outer radii $r_{\rm out}$, and central angular velocities $\Omega_0$.


The actual procedure for generating a specific cloud core with a given value of $\beta$ 
is as follows. First, we choose
the outer cloud core radius $r_{\rm out}$ and find $r_0$ from the condition $r_{\rm out}/r_0=6$.
Then, we find the central surface density $\Sigma_0$ from the relation 
$r_0=\sqrt{A}c_{\rm s}^2/(\pi G \Sigma_0)$ and determine the resulting cloud core mass
$M_{\rm cl}$ from Equation~(\ref{dens}). Finally, the central angular velocity $\Omega_0$
is found from the condition $\beta=0.9\Omega_0^2 r_0^2/c_{\rm s}^2$.
In total, we have simulated numerically the time evolution of 29 cores spanning
a range of initial masses between $0.085~M_\odot$ and $1.8~M_\odot$. 
The adopted initial core mass function is similar to that presented in figure~1 of \citet{Vor10a}. 
The effect of different initial $\Sigma$ and $\Omega$ configurations are considered
in Section~\ref{initcond}.

\section{Classification scheme and disk-to-envelope transition boundary}
\label{scheme}
Modern classification schemes of young stellar objects (YSOs) are designed to distinguish between 
main physical phases of the early stellar evolution
and the spectral energy distribution is often used
to relate a YSO to a particular class \citep[see][for a thorough review]{Evans09}.
In our case, it is more convenient to use the classification breakdown suggested by \citet{Andre93}
and based on the mass remaining in the envelope
\begin{equation}
\begin{array}{ll}
\mathrm{Class~0} \,\,\, & M_{\rm env} \ge 0.5 M_{\rm cl}, \\
\mathrm{Class~I} \,\,\, & 0.1 M_{\rm cl} \le M_{\rm env} < 0.5 M_{\rm cl}, \\
\mathrm{Class~II} \,\,\, & M_{\rm env} < 0.1 M_{\rm cl}.
\end{array}
\label{Andrescheme}
\end{equation}
According to this scheme (hereafter, AWTB scheme), transition between Class 0 and Class I objects 
occurs when the envelope mass $M_{\rm env}$ decreases to half of the initial cloud core 
mass $M_{\rm cl}$. The Class II phase ensues by the time
when the infalling envelope nearly clears and its total mass drops
below 10\% of the initial cloud core mass. Of course, we acknowledge that these numbers
are somewhat arbitrary and the use of other classification diagnostics may introduce
a systematic bias in our results. 

In order to use the AWTB scheme, we need to know which
part of our numerical grid is occupied by the disk and which part belongs to
the infalling envelope at any time instance of the evolution (note that due to the use of the sink
cell we know the stellar mass). This is not trivial and we describe the 
adopted method in detail below.
In order to distinguish between the infalling envelope and burgeoning 
disk, we make use of the gas surface density distribution in the computational grid. 
We set a threshold density for transition between
the disk and envelope at $\Sigma_{\rm d2e}=0.1$~g~cm$^{-2}$, motivated by the observational
fact that protoplanetary disks often decline exponentially with radius at $\Sigma<0.1$~g~cm$^{-2}$
\citep[e.g.][]{Andrews09}. Numerical hydrodynamics simulations of disk structure and evolution
seem to confirm this phenomenon \citep{Vor10b}.

Of course, using only $\Sigma_{\rm d2e}$ is not sufficient and some {\it physically} motivated 
quantity is to be invoked in order to minimize possible uncertainties. Therefore, we also make use of
the gas radial velocity $v_{\rm r}$. The envelope is freely falling onto the disk and
we determine the radial position where this free-fall
motion  terminates near the disk outer edge.

\begin{figure}
  \resizebox{\hsize}{!}{\includegraphics{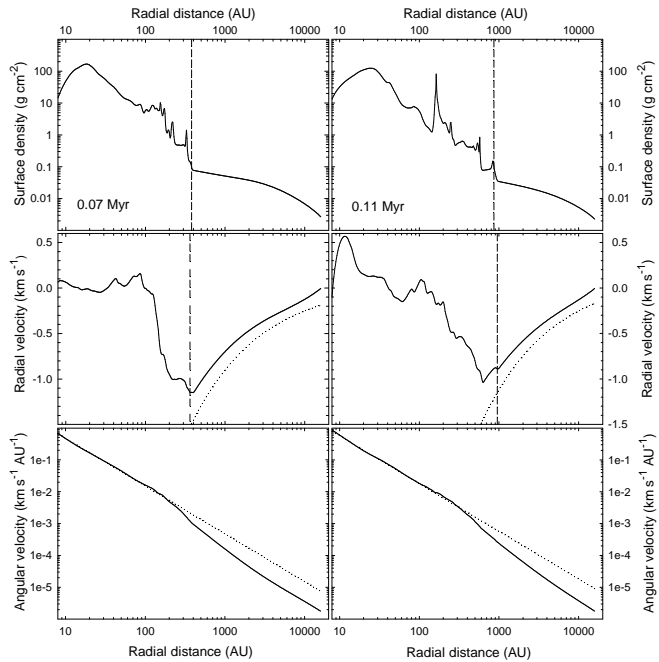}}
      \caption{Radial profiles of the azimuthally averaged gas surface density (top panels),
      radial velocity (middle panels), and angular velocity (bottom panels) at $t=0.07$~Myr (left column)
      and $t=0.11$~Myr (right column) after the formation of the central star. The vertical dashed lines
      mark the position of a tentative surface density threshold for disk to envelope transition 
      $\Sigma=0.1$~g~cm$^{-2}$ (top panels) and the location of the disk outer edge as signalized by
      the characteristic deviation of the radial velocity from that
      expected for a free fall motion (dotted lines in middle panels).
      The dotted line in the bottom panels illustrates a Keplerian rotation. }
         \label{fig1}
\end{figure}

To illustrate our method, we consider a typical model core with $M_{\rm cl}=1.23~M_\odot$,
$\Sigma_0=4.5\times 10^{-2}$~g~cm$^{-2}$, $r_0=2750$~AU, $\Omega_0=1.25$~km~s$^{-1}$~pc$^{-1}$,
and $r_{\rm out}=0.08$~pc.
Figure~\ref{fig1} shows (from top to bottom) the azimuthally averaged radial profiles of 
$\Sigma$, $v_{\rm r}$, and $\Omega$ at $t=0.07$~Myr (left
column) and $t=0.11$~Myr (right column) after the formation of the central star.
The vertical dashed lines indicate an adopted transitional gas surface 
density of $\Sigma_{\rm d2e}=0.1$~g~cm$^{-2}$ (top panels) and the radial position
where the infall motion of gas terminates at the disk outer edge (middle panel). The latter
effect is manifested by a sharp deviation of the radial velocity
from the typical free-fall profile $v_{\rm r}\propto r^{-0.5}$ plotted by the dotted line.
It is seen that the radial locations marked by the vertical dashed lines differ insignificantly 
from each other. 
 It is worth noting that using Keplerian rotation arguments may not be that fruitful
as one might expect. Indeed, the bottom panel demonstrates that a transition from Keplerian
rotation, shown by the dotted line and typical for the disk, to sub-Keplerian one, typical
for the infalling envelope, proceeds smoothly with a fairly wide transitional region. 
The use of this criterion may introduce significant uncertainties in the definition of the disk
radius. 

To summarize, our procedure consists of the following steps. We march from the outer computational boundary
toward the inner one and identify the radial location at which the azimuthally averaged gas surface
density is greater than $\Sigma_{\rm d2e}$. Simultaneously, we determine the location where 
the envelope material hits the disk outer edge. We take the
minimum of these two values as the location of the disk outer edge. To the best of our knowledge, 
this method for discriminating between the disk and envelope is found to be most accurate for our
numerical model (thin disk).
One may argue that we need not to use $\Sigma_{\rm d2e}$ at all, given that the other 
criterion is physically motivated and sufficient. We, however, honestly believe that the exponential
decline of the gas surface density at large radii, as observed in circumstellar disks 
\citep{Andrews09}, is not accidental but physically motivated by, e.g., photoevaporation 
\citep{Hollenbach00}, tidal stripping of the disk outer regions due to 
close encounters between the members of a stellar cluster \citep{Bate03}, or viscous dispersal 
\citep{VB09b}. It is therefore important to impose some cut-off value in $\Sigma$ to take 
these effects into account. We acknowledge
that our choice of $\Sigma_{\rm d2e}=0.1$~g~cm$^{-2}$ may somewhat affect the resulting disk
masses and radii and
we consider the effect of varying $\Sigma_{\rm d2e}$ in more detail in Section~\ref{disk2env}.

Finally, we note that some of our model cores with high enough mass and angular momentum
form binary/multiple systems via disk fragmentation. Whenever this happens, we have to
determine if the companion is still embedded in the disk of the primary or it has detached from 
the natal disk and assembled a circumstellar disk of its own. We do this by scanning 
the azimuthally averaged gas surface density and velocity profiles in our computational domain. 
The companion's position is usually marked by a sharp peak in $\Sigma$ and local flow of
gas pointing to the companion. 
We determine if these two conditions are satisfied. In addition, we postulate that the companion
completely detaches from its natal disk if a gap 
develops between the natal disk and the companion with 
azimuthally averaged $\Sigma<0.01$~g~cm$^{-2}$. According to our tests, this procedure produces
reliable results. When the companion forms and detaches from the natal disk, the mass of the latter
drops accordingly. This process can be seen in the upper-right panel of Figure~\ref{fig8} 
in Section~\ref{initcond}.

\section{Integrated disk properties}
\label{diskproperties}

\begin{figure}
  \resizebox{\hsize}{!}{\includegraphics{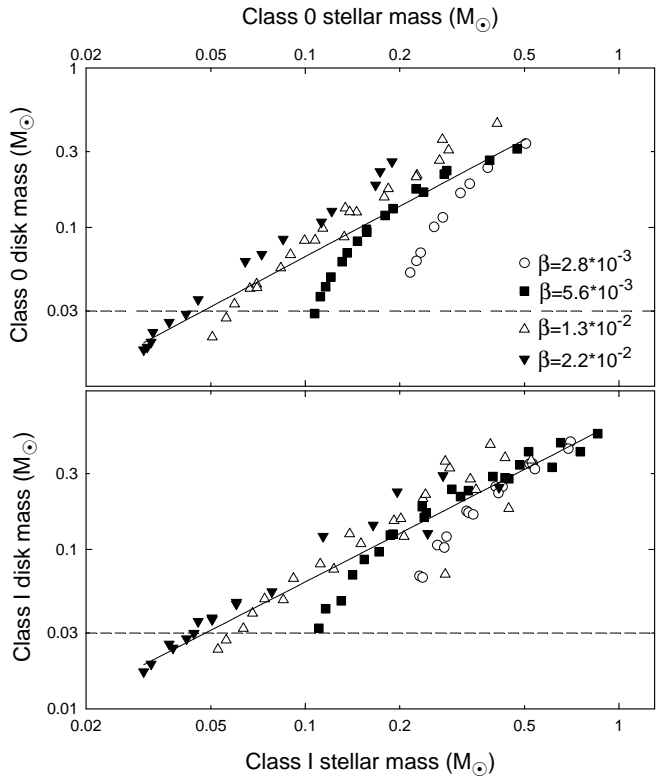}}
      \caption{Disk masses versus stellar masses for the Class~0 objects (top) and Class~I objects (bottom)
      in model set~1 (open circles), model set~2 (filled squares), model set~3 (open triangles up),
      and model set~4 (filled triangles down). The solid lines show the least-squares best fits to the
      model data.}
         \label{fig2}
\end{figure}

\subsection{Disk masses}
\label{diskmass}
A systematic numerical study of disk masses around stars of (sub-)solar mass was done 
by \citet{Vor09a} in the context of polytropic disks.
In this paper, we have implemented a more accurate prescription for the disk thermal physics and 
expanded our parameter space by including cloud cores with higher initial rotational energies. 
Another important improvement over our previous work
is the realization that both disk and stellar masses increase significantly during 
the EPSF. Therefore, presenting the $M_{\rm d}$--$M_\ast$ relation in terms 
of the time-averaged quantities, as was done in \citet{Vor09a}, may not be adequate. 
In this study, we have chosen six typical disk and stellar masses based on the ratio 
of the envelope to cloud core mass, $M_{\rm env}/M_{\rm cl}$. This ratio is useful because 
it gradually decreases with time and can be used as a tracer of the early evolution 
of a protostar. In particular, disk and stellar masses at $M_{\rm env}/M_{\rm cl}$=0.85, 
0.7, and 0.51 are meant to represent typical ones 
in the beginning, in the midway, and at the end of the Class~0 phase, respectively.
For the typical disk and stellar masses in the Class~I phase, we have chosen three values based on 
$M_{\rm env}/M_{\rm cl}$=0.49, 0.3, and 0.1. Of course, these six disk and stellar masses 
cannot represent the whole possible spectrum of masses that can be observationally
detected in the embedded phase of star formation. Nevertheless, they can help to reduce
a possible bias toward a particular evolution stage and can help to make our
disk mass versus stellar mass relation observationally meaningful. 
Finally, we have taken into account the mass of gas contained in the sink cell.
The inner inflow computational boundary is located at $r_{\rm sc}$=6~AU. The gas that
flows through this boundary is assumed to land onto the central star and inner
circumstellar disk with radius $r_{\rm sc}$. The mass partition between these two
objects can affect the obtained stellar and disk masses and is estimated by 
extrapolating the azimuthally averaged
gas surface density profile into the inner 6~AU.
Unfortunately, this procedure was not originally incorporated into the code, but we
perform postprocessing our data to calculate the mass contained in the inner 6~AU.
On average, the estimated mass partition between the star and the inner disk is 10:1,
with the effect of lowering the stellar mass by about $10\%$ and increasing the total disk mass
by the corresponding amount.

Figure~\ref{fig2} presents our model disk masses versus stellar masses in the Class~0 phase (top) and
Class~I phase (bottom) for four model sets, the parameters of which are summarized in Table~\ref{table1}.
In particular, model set~1 is plotted by open circles, model set~2---by filled squares, model set~3---by
open triangles up, and model set~4---by filled triangles down. 
Each symbol of same shape within a given set of models represents an individual 
object formed from a core of distinct mass, rotation rate, and outer radius. Our modeling
covers a wide range of central object masses, staring from sub-stellar objects with $M_\ast\sim 0.03~M_\sun$
and ending with solar-type stars.

Several interesting features can be seen in Figure~\ref{fig2}. First, there is an obvious correlation
between the disk and stellar masses in the EPSF. The least-squares  best fits to the Classes~0
and I data (solid lines) yield the following relations
\begin{eqnarray}
\label{eq1}
M_{\rm d,C0}&=&\left( 0.73^{+0.11}_{-0.09} \right) \, M_{\rm \ast,C0}^{1.05\pm 0.07}, \\
\label{eq2}
M_{\rm d,CI}&=&\left( 0.65^{+0.04}_{-0.05} \right) \, M_{\rm \ast,CI}^{1.0\pm 0.04}, 
\end{eqnarray}
where the subscripts C0 and CI refer to the Class~0 and Class~I phases, respectively, and disk and stellar
masses are in solar masses. It is evident
that the correlations are slightly super-linear and do not depend significantly on the particular phase.
This result is in some dissonance with what was found earlier in the context of polytropic disks \citep{Vor09a},
where a steepening of the $M_{\rm d}$--$M_\ast$ correlation was reported along the Classes~0--II
evolutionary sequence. We believe that this steepening was partly caused by time averaging of the 
corresponding disk and stellar masses over the duration of each phase, which effectively reduced
the range of disk and stellar masses available for constructing the best fits.
It may also be partly caused by a smaller parameter space in our earlier study, which span 
a narrower range of $\beta=(1.2-3.4)\times 10^{-3}$. 

Observations of protostellar disks in the EPSF are extremely difficult due to the obscuration of
light by surrounding envelopes and also due to uncertainties in dust opacities, optical depths, and
disk structure. Disk masses inferred using complicated disk plus envelope models that
employ Monte-Carlo radiative transfer codes and fit various circumstellar dust distributions to 
the measured spectral energy distribution and 
(sub-)millimeter continuum emission exist only for a handful of objects 
\citep[e.g.][]{Brown00,Andrews05,Eisner05,Jorgensen09}. 
 On the other hand, the T~Tauri phase lacks notable envelopes and disk masses in this phase
are inferred for a number of objects. Yet, there is conflicting evidence as to 
the relation between disk and stellar masses in the T~Tauri phase. For instance, \citet{Natta01} 
found a marginal correlation between $M_{\rm d}$ and $M_\ast$, 
albeit with a substantial dispersion. On the other hand, \citet{Andrews05} and \citet{Eisner08}
claimed no correlation. If we assume by extrapolation the existence of a similar 
(to Equations~(\ref{eq1}) and (\ref{eq2})) correlation between $M_{\rm d}$ 
and $M_\ast$ in the T~Tauri phase, then our theoretical predictions are more in line with what was
found by \citet{Natta01}. 

The numerically obtained near-linear correlation between disk and stellar masses can be
broken if a large population of objects occupies both the upper
left and lower right portions of the $M_{\rm d}$--$M_\ast$ phase space in Figure~\ref{fig2}. 
While it is feasible that there exists a fraction of stars
with disk-to-star mass ratios low enough to populate the lower-right corner
of Figure~\ref{fig2} (but see discussion in Section~\ref{ratios}), it is unlikely that 
an equal fraction of 
sub-stellar objects with disks considerably more massive than the star 
could populate the upper-left corner of Figure~\ref{fig2}.
The existence of such systems finds little observational \cite[see e.g.][]{Andrews05} and theoretical
support \citep[see e.g.][]{Kratter10}. As demonstrated by \citet{Vor10a}, systems with
equal disk and stellar masses are short-lived and quickly evolve into binary/multiple
systems with disk masses considerably smaller than those of the stars.
We conclude the our $M_{\rm d}$--$M_\ast$ correlation may weaken somewhat
if more model cores with low values of $\beta$ are considered (such cores are expected to form low-mass
disks due to small centrifugal radii), but it cannot vanish completely.

In accordance with our earlier results, we see a wide scatter in the derived disk masses for stars of
equal mass, in particular for those in the intermediate mass range 0.05--0.5~$M_\sun$. This scatter
is caused by the fact that cloud cores with higher rotation rates (as defined by $\beta$ in our models)
tend to form more massive disks. This implies that the distribution in initial conditions of cloud cores
results in a scatter of disk masses.

\begin{table*}
\begin{center}
\caption{Averaged disk properties}
\label{table3}
\begin{tabular}{ccccccccc}
\hline\hline
Phase & $M_{\rm d}^{\rm mean}$ & $M_{\rm d}^{\rm mdn}$ & $M_{\rm d}^{\rm max}$ & $\xi^{\rm mean}$ 
& $\xi^{\rm mdn}$ & $\xi^{\rm max}$  & $r_{\rm d}^{\rm mean}$ & $r_{\rm d}^{\rm mdn}$  \\
\hline
Class 0  & 0.12 & 0.09 & 0.45 & 0.71 & 0.67 & 1.30 & 230 & 140  \\
Class I  & 0.18 & 0.15 & 0.53 & 0.66 & 0.66 & 1.18 & 480 & 290  \\
 \hline
\end{tabular} 
\tablecomments{Disk masses are in $M_\sun$ and radii are in AU.}
\end{center}
\end{table*} 

Table~\ref{table3} presents our model mean, median, and maximum disk masses 
($M_{\rm d}^{\rm mean}$, 
$M_{\rm d}^{\rm mdn}$, and $M_{\rm d}^{\rm max}$, respectively) calculated from the data shown 
in Figure~\ref{fig2}. Our 
derived values are $M_{\rm d,C0}^{\rm mean}=0.12~M_\sun$
and $M_{\rm d,CO}^{\rm mdn}=0.09~M_\sun$ in the Class~0 phase and also $M_{\rm d,CI}^{\rm mean}=0.18~M_\sun$
and $M_{\rm d,CI}^{\rm mdn}=0.15~M_\sun$ in the Class~I phase. 
The existing estimates seem
to yield {\it lower} mean disk masses, albeit with a considerable scatter.
For instance, \citet{Andrews05} reported $M_{\rm
d,CI}^{\rm mean}=0.03~M_\sun$ for Class~I sources in the
Taurus-Auriga star formation region, while
\citet{Brown00} found $M_{\rm d,C0}^{\rm mean}=0.01~M_\sun$ for Class~0 objects in the Perseus and Serpens
molecular clouds. A recent sub-millimeter survey of low-mass protostars in the Class 0 and I 
phases by \citet{Jorgensen09} revealed a somewhat higher mean disk mass of 0.05~$M_\sun$, yet a factor
of 2--3 lower than our mean values.  
The same tendency of observationally inferred disk masses being smaller than those obtained from numerical
simulations seems to extend to the later phases of stellar evolution \citep{Vor09a}.

\begin{figure}
  \resizebox{\hsize}{!}{\includegraphics{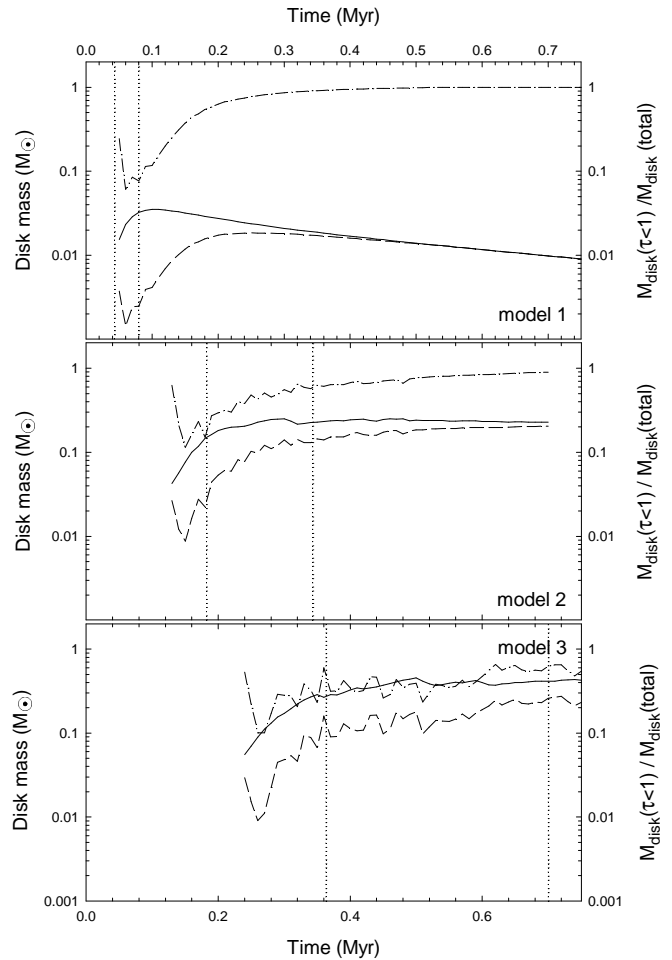}}
      \caption{Time evolution of disk masses in model~1 (top), model~2 (middle), and model~3 (bottom).
      In particular, the solid lines represent the total disk mass $M_{\rm d}$, while the dashed lines
      show the mass $M_{\rm d}(\tau<1)$ of those disk regions that are characterized by 
      the frequency-integrated optical depth to the midplane $\tau<1$. 
      The dash-dotted lines are the ratio $M_{\rm d}(\tau<1)/M_{\rm d}$. }
         \label{fig3}
\end{figure}

Given the difficulties with the observational determination of disk masses in the EPSF and a low number
statistics of embedded sources, we believe that drawing any firm conclusions from the comparison 
of our numerical results with the observations at this stage may be premature until more 
powerful observational facilities (like ALMA) come into work and more observational data  
become available. Nevertheless, we would like point out one possibility that may 
explain why the observationally inferred disk masses seem to be systematically lower
than our model estimates.
We believe that this discrepancy is at least partly caused by high opacity of 
the inner disk regions, which may contain a sizeable fraction of the total mass budget \citep{Rice10,Vor10a}.
This possibility has also been put forward recently by \citet{Greaves10}.

To estimate the magnitude of the mass deficit
due to high disk opacity, we calculate the disk mass that is characterized 
by the optical depth to the midplane $\tau<1$ and compare the resulting value with 
the total disk mass. 
We consider three typical cores with distinct initial masses: model~1 is characterized 
by $M_{\rm cl}=0.2~M_\sun$, while models~2 and~3 have $M_{\rm cl}=0.85~M_\sun$ and 
$M_{\rm cl}=1.7~M_\sun$, respectively. The ratio $\beta$ is identical for these models 
and is set to $5.6\times 10^{-3}$. Figure~\ref{fig3} shows the total disk 
mass ($M_{\rm d}$, solid lines), the mass of disk regions with $\tau<1$ ($M_{\rm d}(\tau<1)$, 
dashed lines), 
and the ratio $M_{\rm d}(\tau<1)/M_{\rm d}$ (dash-dotted lines) as a function of time since the onset
of gravitational collapse. In particular, the top panel shows the results for model~1 (top), while
the middle and bottom panel present the data for model~2 and 3, respectively. The vertical dotted lines
mark the onset of the Class~I phase (left) and Class~II phase (right). The value of 
$M_{\rm d}(\tau<1)$ is used as a proxy to what can be expected from 
the measurements of disk masses in (partly) optically thick protostellar disks.

As anticipated, $M_{\rm d}(\tau<1)$
is always smaller than $M_{\rm d}$ and the difference is particularly large in the embedded
phase but tends to diminish in the Class II phase. The rate of convergence between $M_{\rm d}(\tau<1)$
and $M_{\rm d}$ is however different in models with distinct core masses. The ratio $M_{\rm d}(\tau<1)/M_{\rm
d}$ approaches unity much slower in higher-$M_{\rm cl}$ models, indicating that massive disks remain
partly optically thick for a longer time. On a more qualitative side, $M_{\rm d}(\tau<1)$ in the 
Class~0 phase may be a factor of 5--10 smaller than the total disk mass. 
In the Class~I phase, the difference is less impressive (usually a factor of 2--3) but may become much
stronger for low-mass disks.  We note that we have excluded a possible contribution of 
the sink cell to the total disk mass budget due to a difficulty with calculating self-consistently
the optical depth there. However, disk regions at $r<r_{\rm sc}=6$~AU are usually optically thick and
this fact should further magnify the effect. 
We conclude that, due to high opacity, the observationally inferred disk masses may be seriously 
underestimated in the Class~0 phase and, to a lesser extent, in the Class~I phase.

Finally, we point out that both modern theories of giant planet formation, those of core accretion and
direct gravitational instability, require that the gas surface density
in circumstellar disks be several times greater than that of the minimum mass solar
nebular \citep{Pollack96,Boss98,Ida04}. 
Taking 350~AU as a characteristic disk radius (see Table~\ref{table3}) and 
adopting the MMSN gas surface density profile from \citet{Hayashi85}, we obtain 
$\approx0.03~M_\odot$ for the mass of the MMSN.
Figure~\ref{fig2} shows that most of our models have disks with
masses in excess of 0.03~$M_\sun$ (horizontal dotted line) 
in both the Class 0 and Class I phases of star formation. 
The least-squares fits indicate that disk masses start to systematically exceed that of the 
MMSN for objects with stellar mass as low as $M_\ast\approx 0.05~M_\odot$.

At a first glance, the abundance of objects with disk mass greater than that of the MMSN
may suggest 
that giant planets may start forming as early as in the Class 0 
and I phases of stellar evolution, a supposition also advocated by \citet{Greaves10}
based on the statistics of observationally inferred disk masses in different stages of star formation.
This may certainly be so for the core accretion mechanism.
However, the direct gravitational instability scenario encounters serious difficulties in the EPSF.
Although it is true that protostellar disks in the EPSF are more prone to gravitational instability
and fragmentation than in any other stage of their evolution, it is also true that the conditions
for giant planet {\it survival} are least favorable in the EPSF. As \citet{VB06,VB10b} have demonstrated,
gravitational interaction of protoplanetary embryos with natal spiral arms results in rapid 
inward migration of the embryos to the inner few AU. The future prospects for the embryos are unclear
due to the specifics of modeling (sink cell at a few AU), but judging from fast migration timescales
(a few orbital periods) we believe that most of them will be tidally
destroyed and absorbed by the central star, leading to a luminosity outburst similar to that of 
the FU-Orionis-type or EX-Lupi-type stars. However, if dust sedimentation and H$_2$ 
dissociation timescales are comparable to that of the migration one (for instance, in massive enough
embryos at wide enough orbits), then some of the embryos may survive this migration, 
loose their upper atmospheres 
and form either metal-reach giants or even earth-type planets on orbits of order a few~AU 
\citep{Boley10,Nayakshin10,Cha10}. Alternatively,
embryos that form in the very late EPSF are less exposed to rapid radial migration; they may
open a gap and settle on wide orbits of order 100~AU \citep{VB10a}. However, the expected 
frequency of such systems is 10\% at best.

\subsection{Disk-to-star mass ratios}
\label{ratios}
The disk-to-star mass ratio $\xi=M_{\rm d}/M_\ast$ is another important diagnostic 
that can give us some useful 
information about stability, fragmentation, and mass transport properties in circumstellar disks.
Observationally inferred $\xi$ vary in wide limits, with most of the values lying in the 
0.01--0.05 range \citep[e.g.][]{Beckwith90,Mannings00,Andrews05}. As to a possible dependence of 
disk-to-star mass ratios on stellar masses, this issue is unsettled and some authors claim no 
correlation \citep[e.g.][]{Natta01}, while others
find a negative correlation, i.e., higher $\xi$ for lower $M_\ast$ \citep[e.g.][]{Mannings00,Andrews05}.

\begin{figure}
\resizebox{\hsize}{!}{\includegraphics{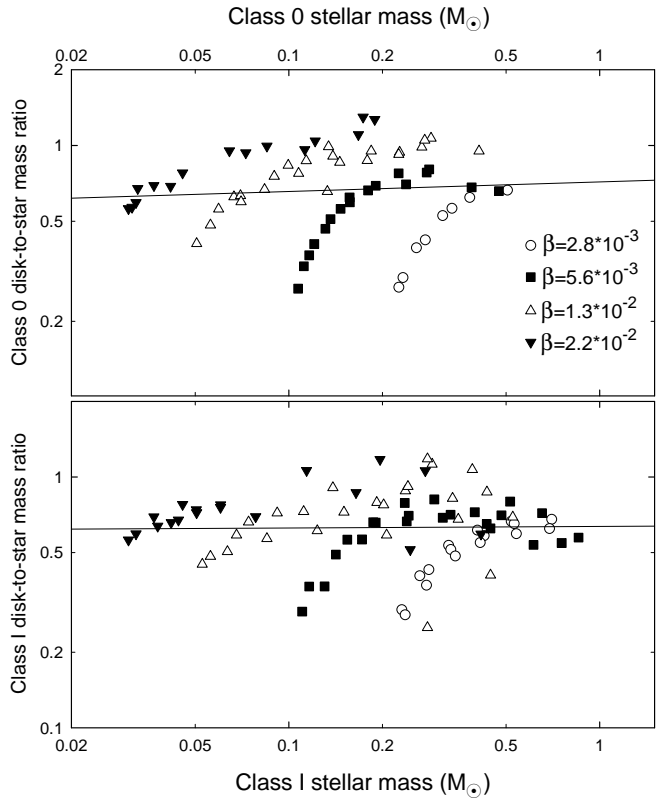}}
      \caption{Disk-to-star  mass ratios versus stellar masses for the Class~0 objects (top) and 
      Class~I objects (bottom)
      in model set~1 (open circles), model set~2 (filled squares), model set~3 (open triangles up),
      and model set~4 (filled triangles down). The solid lines show the least-squares best fits to the
      model data.}
         \label{fig4}
\end{figure}

Figure~\ref{fig4} presents our model disk-to-star mass ratios derived from the data
of Figure~\ref{fig2} for the Class~0 (top) and Class~I (bottom) objects. 
The meaning of the symbols is the same as in Figure~\ref{fig2}.
The mean disk-to-star mass ratios in the Class~0 and I phases 
are $\xi_{\rm C0}^{\rm mean}=0.71$ and $\xi_{\rm CI}^{\rm mean}=0.66$, respectively. 
These values are considerably greater than those inferred from observations 
owing to higher disk masses derived in our modeling. 
In addition, measurements of disk masses are mostly done for the Class~II
objects, for which we could expect lower values of $\xi$ due to accretion of matter from the disk onto
the star, disk viscous dispersal, and photoevaporation.


Another interesting feature seen in Figure~\ref{fig4} is the apparent lack of objects with $\xi<0.2$.
The matter is that our earliest measurements of disk masses are at $M_{\rm
env}/M_{\rm cl}=0.85$. By this time instance, disks have already gained some non-negligible mass owing
to inefficient inward mass transport in the very early phases of disk formation (when gravitational instability is underdeveloped and viscous torques alone cannot cope with
mass loading from the infalling envelope). As a result, the disk quickly grows in mass 
until gravitational
instability and fragmentation set in and decelerate the process of mass growth \citep{Vor09a,Vor10b}.
The net effect is that systems with $\xi<0.2$ are expected to be statistically rare. In addition,
observational estimates of $\xi$ at the very early phases of star formation when $M_{\rm env}/M_{\rm
cl}>0.85$ are very challenging.

The least-squares best fits shown in Figure~\ref{fig4} by the solid lines yield the following 
relations between disk-to-star mass ratios in the Class~0 and Class~I phases 
($\xi_{\rm C0}$ and $\xi_{\rm CI}$, respectively) and the corresponding stellar masses (in $M_\sun$)
\begin{eqnarray}
\xi_{\rm C0}&=&\left( 0.72 \pm 0.1 \right) \, M_{\rm \ast,C0}^{0.04\pm 0.07}, \\
\xi_{\rm CI}&=&\left( 0.64 \pm 0.04  \right) \, M_{\rm \ast,CI}^{0.01\pm 0.04}. 
\end{eqnarray}
It is evident that $\xi$ demonstrates little correlation with $M_\ast$ in the EPSF.
The maximum disk to star mass ratios found in our modeling in the Classes~0 and I phases are 
$\xi_{\rm C0}=1.3$ and $\xi_{\rm CI}=1.18$, respectively. This is in agreement with a recent 
study by \citet{Kratter10} and a factor of 3 greater than was previously reported by \citet{Vor09a}
in the context of polytropic disks.

It is interesting that the maximum $\xi$ is attained by objects with intermediate stellar masses 
$M_\ast\approx0.2-0.3~M_\ast$ rather than by the upper-mass stars in our sample ($M_\ast\approx 1.0~M_\ast$)
as might have been intuitively expected. 
 The matter is that the mass transport in disks around low-mass stars $M_\ast\la0.2~M_\sun$ 
is dominated by viscous torques rather than by gravitational ones \citep{VB09b}. 
As the stellar mass grows, so does the mass of the parent core
and the viscous torques fail to transport a growing amount of infalling core material through 
the disk and onto the star. This causes $\xi$ to increase in the 0.03--0.2~$M_\sun$ mass regime.
For stars with $M_\ast\ga0.2-0.3~M_\ast$, the
situation changes qualitatively---mass transport in their disks becomes dominated by 
gravitational torques and, in even to a greater extent, my quick inward migration of 
forming fragments \citep{VB10b}. Both mechanisms appear to be considerably
more efficient mass transport agents than viscous torques
and the corresponding disk-to-star mass ratios start to decline.
Another factor contributing to the decline in $\xi$ is the formation of 
binary/multiple systems in massive enough disks \citep{Vor10b}.
 

\subsection{Disk radii}
\label{diskradii}


The size of a circumstellar disk is an important characteristics containing indirect information about
the disk stability properties. Numerous numerical and theoretical studies indicate that 
circumstellar disks of larger size are more prone to the development of gravitational 
instability and  fragmentation, in part due to an increased disk mass and in part due to 
faster radiative cooling and lesser stellar irradiation/viscous heating at large radii. 
Indeed, the stellar irradiation flux declines with radius as $h/r^3\propto r^{-1.75}$ for 
a flared 
disk with the typical ratio of the scale height to radius $h/r\propto r^{0.25}$, 
which makes extended disks 
colder (for a fixed stellar luminosity) and hence more prone to gravitational instability. 
At the same time, the ratio 
$t_{\rm cool}/t_{\rm dyn}$ of the characteristic cooling time $t_{\rm cool}=e/\Lambda$ (where
$e$ is the internal energy per surface area and $\Lambda$ is the cooling rate from the disk surface)
to the dynamical time $t_{\rm dyn}=2\pi\Omega^{-1}$ declines with radius as $r^{-3}$ in the 
optically thick regime (for $\Sigma\propto r^{-3/2}$, $T\propto r^{-1/2}$, and spatially 
independent dust opacity) and becomes $r$-independent in the optically thin case, 
again suggesting that
the outer disk regions are more susceptible to gravitational instability.

\begin{figure}
  \resizebox{\hsize}{!}{\includegraphics{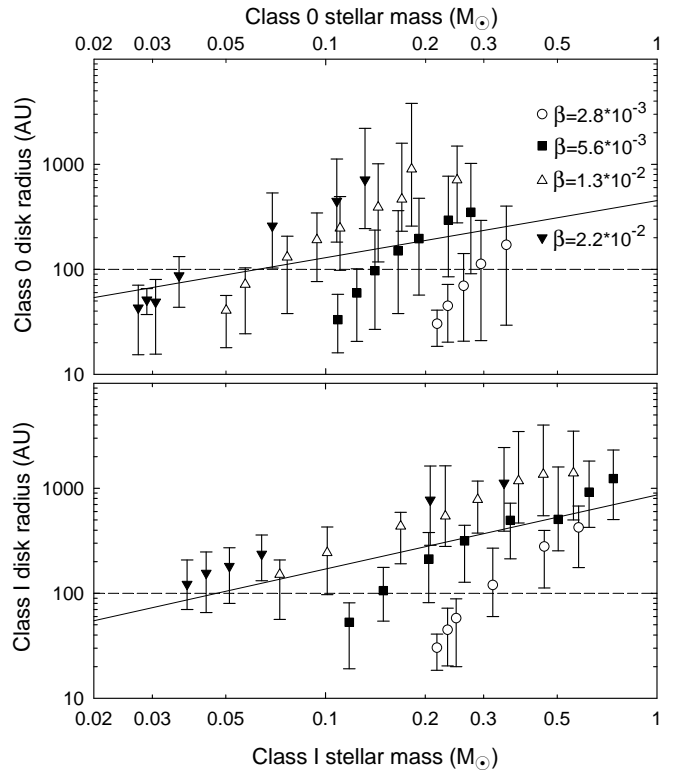}}
      \caption{Time-averaged disk radii versus time-averaged stellar masses for the Class~0 
      objects (top) and Class~I objects (bottom)
      in model set~1 (open circles), model set~2 (filled squares), model set~3 (open triangles up),
      and model set~4 (filled triangles down). Vertical bars represent the minimum and maximum disk
      radii in each model. The solid lines show the least-squares best fits to the
      model data, while the dashed lines mark a fiducial critical radius for disk fragmentation 
      via gravitational instability.}
         \label{fig5}
\end{figure}

Figure~\ref{fig5} presents the time-averaged disk {\it outer} radii $\langle r_{\rm d} \rangle$ 
as a function of the time-averaged stellar masses $\langle M_\ast \rangle$ 
in the Class~0 phase (top) and Class~I phase (bottom). The meaning of the symbols is the same as in
Figures~\ref{fig2} and \ref{fig4}. 
We perform time averaging over the duration 
of the corresponding evolutionary phase in order to smooth out large radial pulsations
seen in most disks during the EPSF \citep[see][for more detail]{Vor10b}. 
In order to inform the reader on the magnitude of these pulsations, we plot vertical bars representing
the minimum and maximum disk radii found in each model. To avoid some ambiguity in the 
definition of the minimum disk radius, we calculate this quantity at the time when the envelope 
mass drops to $0.85\%$ that of the initial cloud core mass, i.e., when $M_{\rm env}/M_{\rm cl}=0.85$.

It is seen that, for a given stellar mass, disk radii may 
vary by as much as two orders of magnitude. The mean and median disk radii in the Class~0 phase 
are $r_{\rm d,C0}^{\rm mean}=230$~AU and $r_{\rm d,C0}^{\rm mdn}=140$~AU, respectively. The corresponding
values in the Class~I phase are $r_{\rm d,CI}^{\rm mean}=480$~AU and $r_{\rm d,CI}^{\rm mdn}=290$~AU.
The characteristic radii in the Class~I phase increase by a factor of 2 as compared to those
in the Class~0 phase, reflecting the ongoing disk growth and expansion due to infall 
of the envelope material during the EPSF.

Figure~\ref{fig5} reveals a mild trend of protostellar disks to grow in size with increasing 
stellar mass (and disk mass due to the near-linear $M_{\rm d}$--$M_\ast$ relation). The least-squares
best fit to the model data yields the following relations between the time-averaged disk radii (in AU)
and time-averaged stellar masses (in $M_\sun$) 
\begin{eqnarray}
\langle r_{\rm d,C0} \rangle&=&\left( 450^{+330}_{-190}  \right) \, \langle M_{\rm \ast,C0}\rangle^{0.55\pm 0.25}, \\
\langle r_{\rm d,CI}\rangle &=&\left( 850^{+380}_{-260}  \right) \, \langle M_{\rm \ast,CI}\rangle^{0.70\pm 0.20},
\end{eqnarray}
where indices C0 and CI refer to the Class~0 and Class I phases, respectively. 
These sub-linear correlations can be understood if we assume that protostellar disks
attain a self-regulatory state in which the disk surface density does not increase notably
with increasing stellar mass. In this case, the mass of such disks 
will grow due to an increase in the disk size. This in turn implies 
$r_{\rm d} \propto M_{\rm d}^{0.5} 
\propto M_\ast^{0.5}$ (for $\xi\equiv M_{\rm d}/M_\ast\propto M_\ast$, 
see Figure~\ref{fig4}). A somewhat steeper dependence found in our numerical simulations 
is explained by the fact that in reality the disk mass grows due to an increase in both the 
disk size and gas surface density, though the former process dominates the latter 
in sufficiently massive disks formed from cloud cores with mass $M_{\rm cl}\ga 0.9~M_\sun$ 
\citep{Vor10b}.

Observational estimates of disk sizes in the EPSF encounter the same difficulties 
as in the case of disk masses and only a few objects have been studied so far.
For instance, \citet{Enoch09} reported a massive disk around Serpens FIRS~1, a well-known 
Class~0 source, with radius of order 300--500~AU. \citet{Hogerheijde01} found an even larger disk 
around L1489~IRS, presumably an object in transition between the embedded and T~Tauri phases, 
with radius of order 2000~AU. More observations of embedded sources are needed to confirm 
that large disks are typical for the embedded phase of star formation.
 
The statistics is wider in the case of Class~II disks.
Observational estimates of disk sizes for $1.0$-Myr-old protoplanetary 
disks in the Orion star-forming region illuminated by the UV radiation of massive 
stars were done by \citet{Vicente05}. They found typical disk radii around a large sample 
of late-G to late-M stars in the Trapezium cluster to be in the 50--200~AU range,
with a median value of 70~AU for protoplanetary disks and 130~AU for a subset of silhouette disks. 
These estimates are considerably smaller than our model values.
On the other hand, \citet{Mann09} reported disk radii for two massive protoplanetary disks 
situated beyond the Trapezium cluster in Orion to be of order 300~AU, suggesting that strong 
UV radiation from OB stars may have evaporated part of the nearby disks and truncated their radii.
This view is supported by the estimates of disk radii in the Taurus and Ophiuchus star forming regions,
less extreme than that of the Orion, performed
by \citet{Andrews07}.  They found disk radii to lie in a wide range between 50~AU and 1000~AU,
with a median value of 200~AU. This value is in better agreement with our median values, which
are $r_{\rm d,C0}^{\rm mdn}$=140~AU for the Class~0 disks and $r_{\rm d,CI}^{\rm mdn}$=290~AU 
for the Class~I disks.  We stress, however, that the observationally inferred disk radii really correspond
to a sensitivity limit to the surface brightness profile of the continuum emission and real disks may
be somewhat larger.

Many theoretical and numerical studies indicate that disk fragmentation is unlikely in the inner 50--100~AU due to insufficient cooling and elevated viscous and stellar irradiation heating \citep{Rafikov09,
Clarke09,VB10b}. In the context of planet formation, this means that
disks must be greater than 100~AU in radius in order to form giant planets via disk fragmentation. 
Figure~\ref{fig5} reveals a substantial fraction of objects with time-averaged disk radii 
exceeding 100~AU (as marked by the horizontal dashed lines), in the Class~I phase in particular.
These objects are formed from prestellar cores of sufficiently high initial mass and angular momentum.
The least-squares fits indicate that disk radii start to exceed 100~AU for objects with stellar 
mass as low as 0.05--0.07~$M_\odot$.

\subsection{Characteristic gas surface densities and midplane temperatures}

In this section, we provide typical gas surface densities and temperatures of protostellar disks
in the embedded phase of their evolution. Due to the use of the sink cell, we can calculate typical
values only at relatively large radii. This information is of particular interest for the 
giant planet formation mechanism via direct gravitational instability, which has been predicted and
demonstrated to 
operate in the disk outer regions only \citep[e.g.][]{Dodson09,Rafikov09,Clarke09,VB10a,Boley10}.

\begin{figure}
  \resizebox{\hsize}{!}{\includegraphics{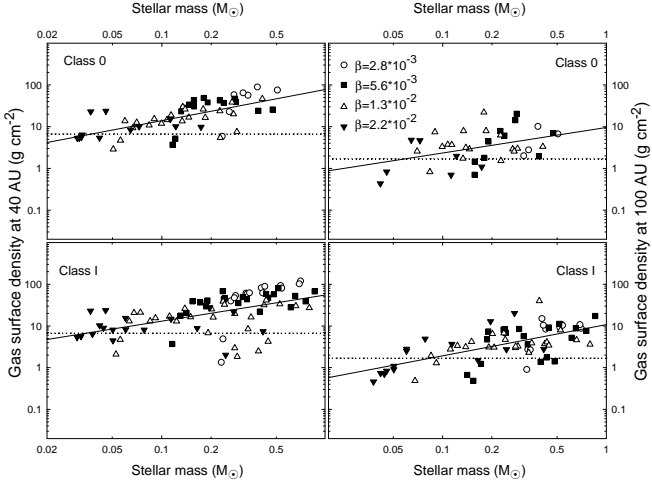}}
      \caption{Typical disk surface densities at 40~AU (left column) and 100~AU (right column) in the
      Class~0 phase (top row) and Class~I phase (bottom row). The solid lines provide the least-squares
      fits to the model data, while horizontal dotted lines mark the gas surface densities in the MMSN
      at 40~AU and 100~AU.}
         \label{fig6}
\end{figure}

Figure~\ref{fig6} presents gas surface densities at 40~AU ($\Sigma_{40}$, left column) 
and 100~AU ($\Sigma_{100}$, right column) from the star as a function of stellar mass $M_\ast$ 
for the same four sets of model cores as in Figure~\ref{fig2}. 
In particular, the top and bottom panels correspond to the data in the Class~0 and Class~I phases, 
respectively. As in Figure~\ref{fig2}, the gas surface densities and stellar masses are calculated 
at the beginning, in the midway, and at the end of each evolutionary phase. 
Because we are interested in {\it disk} surface densities, we have excluded model cores
that fail to form large enough disks at the evolutionary times of interest 
(about 10\% of the total number of models).

The least squares best fits (solid lines) yield the following relations
\begin{equation}
\Sigma_{\rm 40,C0} = 80^{+20}_{-15} \, M_{\ast,C0}^{0.75\pm0.1},
\end{equation}
\begin{equation}
\Sigma_{\rm 40,CI} = 9.5^{+4.5}_{-3.5}\, M_{\ast,CI}^{0.6\pm0.2},
\end{equation}
\begin{equation}
\Sigma_{\rm 100,C0} = 55^{+12}_{-10}\, M_{\ast,C0}^{0.6\pm0.1},
\end{equation}
\begin{equation}
\Sigma_{\rm 100,CI} = 10\pm 2\, M_{\ast,CI}^{0.75\pm0.1},
\end{equation}
where 
gas surface densities and stellar masses are in g~cm$^{-2}$ and solar masses. 
It is seen that the correlations between $\Sigma$ and $M_\ast$ are sub-linear, 
more massive disks are generally denser than their low-mass counterparts,
and $\Sigma$ declines with radius as found in many theoretical and numerical studies.

As discussed in Section~\ref{diskmass},
the gas surface density should be several times greater than that of the minimum mass 
solar nebular (MMSN) for the gas giants to form either through core 
accretion or gravitational instability. The horizontal dashed lines in Figure~\ref{fig6} 
mark the gas surface densities of the MMSN at 40~AU and 100~AU as calculated from the radial 
profile of \citet{Hayashi85}: $\Sigma_{\rm MMSN}[\mathrm{g~cm^{-2}}]=1700 (r/1~\mathrm{AU})^{-1.5}$.
The gas surface density at 100~AU is most relevant to giant planet formation 
via direct gravitational instability,
which is supposed to operate at distances of order 100~AU and beyond.
The least-squares fits indicate that gas surface densities at 100~AU 
start to systematically exceed that of the MMSN for objects with stellar mass as low as 
0.07--0.09~$M_\odot$.

\begin{figure}
  \resizebox{\hsize}{!}{\includegraphics{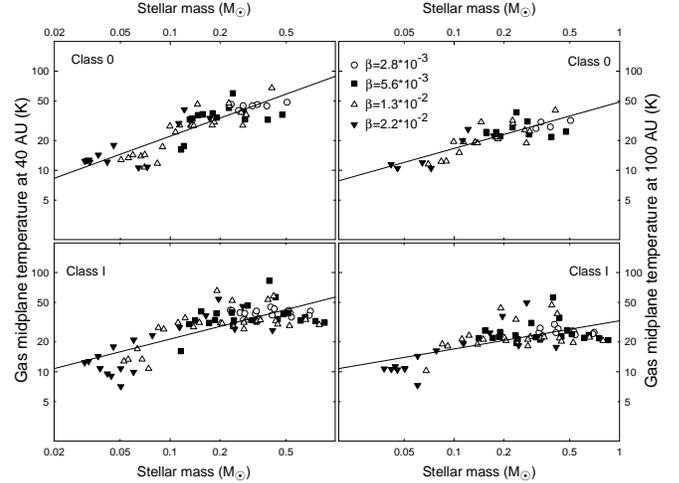}}
      \caption{Typical gas midplane temperatures at 40~AU (left column) and 100~AU 
      (right column) in the
      Class~0 phase (top row) and Class~I phase (bottom row). The solid lines provide the least-squares
      fits to the model data.}
         \label{fig7}
\end{figure}

On the other hand, the core accretion scenario is believed to operate in the inner 10~AU. 
We cannot provide reliable data for these inner regions due to the specifics of our 
modeling. However, we note that as one goes from 100~AU to 40~AU 
the corresponding gas surface densities start to exceed that of the MMSN for objects of 
even smaller stellar mass 
$M_\ast$=0.03--0.05~$M_\odot$ (in contrast to 0.07--0.09~$M_\odot$ for $\Sigma_{\rm 100}$).
If we assume by extrapolation 
that this trend continues to even smaller radii, than planet formation via core accretion 
can also occur in objects with stellar mass as low as 0.03--0.05~$M_\odot$.



Finally, in Figure~\ref{fig7} we provide typical gas midplane temperatures at 40~AU ($T_{40}$, 
left column) and 100~AU ($T_{100}$, right column) for the same models as in Figure~\ref{fig6}.
The top/bottom rows show data for the Class~0/Class~I phases, respectively.
The least squares best fits (solid lines) yield the following relations
\begin{equation}
T_{\rm 40,C0} = 90\pm 8 \, M_\ast^{0.6\pm0.05},
\end{equation}
\begin{equation}
T_{\rm 40,CI} = 50\pm 5\, M_\ast^{0.5\pm0.05},
\end{equation}
\begin{equation}
T_{\rm 100,C0} = 55\pm 4\, M_\ast^{0.45\pm0.04},
\end{equation}
\begin{equation}
T_{\rm 100,CI} = 30 \pm 5\, M_\ast^{0.3\pm0.05}.
\end{equation}
It is seen that massive disks (i.e., those around stars of greater mass) are generally 
hotter than their low-mass counterparts.
On the other hand, the correlations between $T$ and $M_\ast$
are notably weaker than those between $\Sigma$ and $M_\ast$, 
indicating that efficient cooling operates at radii of order 50--100~AU.
There seem to be a slight decrease in the exponents
along the Class~0--Class~I evolutionary sequence. 
One interesting feature of Figure~\ref{fig7} is a notable jump in the gas midplane temperature 
seen at around $M_\ast=0.05-0.1~M_\sun$,
which represents a transition from optically thin to optically 
thick disks.

\section{Model caveats}
\label{caveats}

\subsection{Initial conditions} 
\label{initcond}

\begin{figure}
  \resizebox{\hsize}{!}{\includegraphics{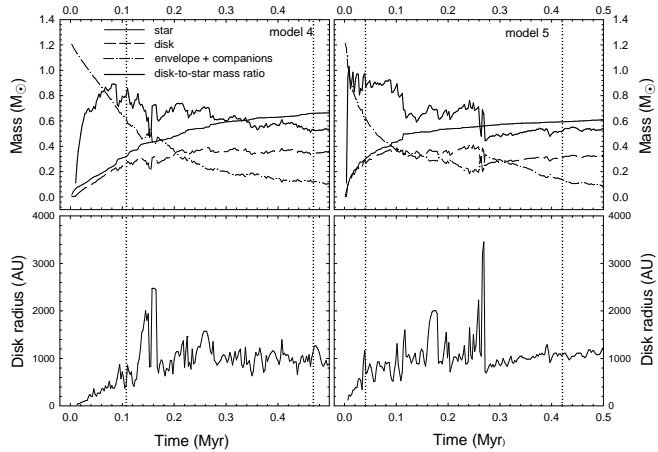}}
      \caption{{\bf Top.} Disk masses (dashed lines), stellar masses (thin solid lines), envelope 
      plus companions masses (dash-dotted lines), and
      disk-to-star mass ratios (thick solid lines) versus time passed since 
      the formation of the central star in model 4 (left) and model~5 (right). {\bf Bottom.}
      Disk radii versus time in model~4 (left) and model~5 (right). The vertical dotted lines mark the
      onset of the Class~I phase (left) and Class~II phase (right).}
         \label{fig8}
\end{figure}
Our model cloud cores have distinct masses (correlated
with variations in $\Sigma_0$ and external pressure) and rotational energies (correlated with
$\Omega_0$). At the same time, the shape of the radial gas surface density and angular velocity 
profiles remains fixed (Equations~(\ref{dens}) and (\ref{omega})).
Although the $\Sigma\propto r^{-1}$ profile is appropriate for 
supercritical cores formed via ambipolar diffusion and, perhaps, for a wider family of cores 
formed via slow gravitational contraction \citep{Dapp09}, it is not guaranteed, however, that this
profile is universal. In particular, the current
model assumes each star is roughly the Jeans mass in its natal environment and, as a consequence,
more massive stars come from a lower pressure environment. However, the assumed independence of 
the shape of prestellar cores from the external pressure may break, for massive star 
formation in particular\footnote{We note, however, that our stellar mass spectrum is limited from above by 
stars with mass roughly equal to that of the Sun.}.
The same arguments apply to the adopted angular velocity profile.
As an alternative, we take the approach of \citet{Boss01} and explore an initially isothermal, 
self-gravitating, sheetlike cloud with volume density depending only on distance from the midplane 
$\rho(z)=\rho(0) \mathrm{sech}^2(z/h)$. Such cylinder-like configurations
have constant surface densities and are also solid-body rotators.
In the following, we consider two specific model prestellar cores that are characterized 
by equal masses $M_{\rm
cl}=1.23~M_\odot$ and ratios $\beta=9\times 10^{-3}$ but distinct initial shapes. 
In particular, model~4 has nonuniform $\Sigma$ and
$\Omega$ described by Equations~(\ref{dens}) and (\ref{omega}), with 
$\Sigma_0=4.5\times 10^{-2}$~g~cm$^{-2}$, 
$r_0=2750$~AU, and $\Omega_0=1.25$~km~s$^{-1}$~pc$^{-1}$,
while model~5 has spatially uniform $\Sigma=1.3\times 10^{-2}$~g~cm$^{-2}$ and 
$\Omega=1.25$~km~s$^{-1}$~pc$^{-1}$. In both models, $r_{\rm out}=0.08$~pc.
These two models are meant to represent two limiting cases 
for the initial shapes of prestellar cores in low-mass star forming regions.

The top panels in Figure~\ref{fig8} present stellar masses $M_\ast$ (thin solid line), disk masses $M_{\rm
d}$ (dashed lines), envelope plus companion masses $M_{\rm env}$ (dash-dotted lines), and 
disk-to-star mass ratios $\xi$ (thick solid lines in model~4 (left)
and model~5 (right) as a function of time passed since the formation of the central star. 
The bottom panels show the disk radius in model~4 (left) and model~5 (right). 
The vertical dotted lines mark the onset of Class~I and Class~II phases of star formation.
The two models show distinct time evolution, as expected for gravitationally unstable 
protostellar disks formed from non-identical prestellar cores,
but when time averaged values are considered it turns out that notable
differences take place only in the Class~0 phase. Tables~\ref{table4} and \ref{table5}
present main model characteristics time-averaged over the duration of Class~0 and Class~I phases, 
respectively. From Table~\ref{table4} it is evident that the nonuniform model~5 has greater 
$\xi_{C0}$, $M_{\rm d,C0}$, $M_{\ast,C0}$, and
$r_{\rm d,C0}$ but smaller $M_{\rm env,C0}$ than the uniform model~4, indicating that protostellar cores with 
spatially uniform profiles of  $\Sigma$ and $\Omega$ sustain higher infall rates onto the disk and
deplete their material faster than cores with spatially declining profiles.
At the same time, Table~\ref{table5} reveals little difference between the two models in the Class~I
phase. We conclude that, if the whole spectrum of initial conditions were taken into account, we might
expect to obtain somewhat higher disk masses, stellar masses, disk-to-star mass ratios, and disk
radii in the Class~0
phase but the corresponding values in the Class~I phase would remain largely unaffected.

\begin{table}
\caption{Time-averaged model characteristics in Class~0 phase}
\label{table4}
\begin{tabular}{cccccc}
\hline\hline
Model & $\langle \xi_{\rm C0} \rangle$ & $\langle M_{\rm d,C0}\rangle$ & $\langle M_{\ast,C0}\rangle$
 & $\langle M_{\rm env,C0}\rangle$  & $\langle r_{\rm d,C0} \rangle$ \\
\hline
 4 & 0.71 & 0.13 & 0.18  & 0.9  &  325 \\
 5 & 0.87 & 0.17 & 0.19  & 0.84 &  453 \\
 \hline
\end{tabular} 
\tablecomments{All masses are in $M_\sun$ and disk radii---in AU.}
\end{table} 

\begin{table}
\caption{Time-averaged model characteristics in Class~I phase}
\label{table5}
\begin{tabular}{cccccc}
\hline\hline
Model & $\langle \xi_{\rm CI} \rangle$ & $\langle M_{\rm d,CI}\rangle$ & $\langle M_{\ast,CI}\rangle$
 & $\langle M_{\rm env,CI}\rangle$  & $\langle r_{\rm d,CI} \rangle$ \\
\hline
 4 & 0.63 & 0.34 & 0.55  & 0.27 &  1065 \\
 5 & 0.64 & 0.33 & 0.52  & 0.29 &  1060 \\
 \hline
\end{tabular} 
\tablecomments{All masses are in $M_\sun$ and disk radii---in AU.}
\end{table} 

\subsection{Discriminating between the disk and infalling envelope}
\label{disk2env}
\begin{figure}
  \resizebox{\hsize}{!}{\includegraphics{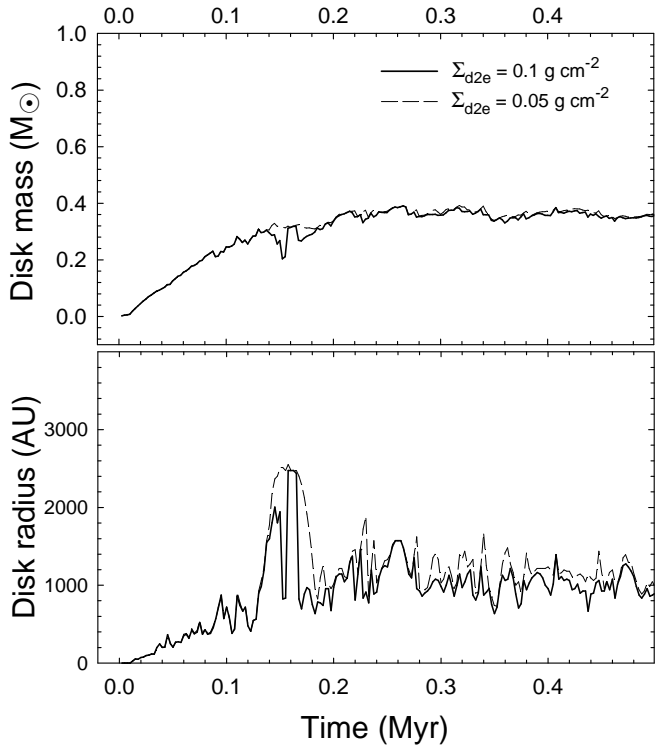}}
      \caption{{\bf Top.} Disk mass versus time for $\Sigma_{\rm d2e}=0.1$~g~cm$^{-2}$ (solid line)
      and $\Sigma_{\rm d2e}=0.05$~g~cm$^{-2}$ (dashed line). {\bf Bottom.} Disk radii for the corresponding
      vales of $\Sigma_{\rm d2e}$. }
         \label{fig9}
\end{figure}

Throughout the paper, we have adopted a characteristic transitional density between the disk
and envelope of $\Sigma_{\rm d2e}=0.1$~g~cm$^{-2}$. Here, we illustrate the effect of a factor
of two smaller $\Sigma_{\rm d2e}=0.05$~g~cm$^{-2}$. Top panel in Figure~\ref{fig9} shows the
disk mass versus time for $\Sigma_{\rm d2e}=0.1$~g~cm$^{-2}$ (solid line) and $\Sigma_{\rm d2e}=0.05$~g~cm$^{-2}$
(dashed line), while the bottom panel presents disk radii for the corresponding transitional 
densities. It is evident that the disk mass is little affected by a factor of 2 drop in $\Sigma_{\rm
d2e}$, owing to an exponentially declining gas surface density in the disk outer regions. The disk radii are somewhat
more sensitive to $\Sigma_{\rm d2e}$, yet showing only a factor of 1.2 increase on average.
Figure~\ref{fig9} convincingly demonstrates that our model results are not critically sensitive to 
the choice of $\Sigma_{\rm d2e}$.

\subsection{Jet efficiency and limitations of the thin-disk approximation}
In this paper, we have taken the simplest approach and assumed that 10\% of the mass passing
through the sink cell is subsequently ejected with protostellar jets. This number has some 
motivation since observations of the variation of jet diameters with distance
from their driving sources have been consistent with models giving the jet efficiency of $>3\%$ \citep{Ray07}.
At the same time, theoretical and numerical studies suggest a higher jet efficiency, 
of up to 70\% depending on the degree of core flattening \citep[e.g.][]{Shu94,Matzner00}, due to a three-dimensional nature of the 
jet phenomenon. In this case, not only the wind material is lost but also the envelope material is entrained
and evacuated, driving massive molecular outflows.
It is however very difficult
to self-consistently take into account the removal of the envelope material by jets in our 
two-dimensional simulations.
We postpone the solution of this problem to a later time, when an accurate vertical disk structure is
implemented in the code, and note that a higher jet efficiency (than currently adopted) may
result in lower final stellar and disk masses.

Another issue associated with the thin-disk model is that the envelope material is forced to
land all at once onto the disk outer edge, while in three dimensions it may
cover a larger area above and below the disk surface. However, numerical simulations of 
gas trajectories in flaring disks (our model disks have the aspect ratio $h/r\propto r^{0.25}$) indicate that the bulk of the envelope material (up to 80\%)
still lands onto the disk outer edge \citep{Visser09}.
The remaining fraction falls onto the disk surface, triggering the streaming instability due
to the vertical sheer in the gas velocity. This instability acts to further destabilize the disk and
increase the mass and angular momentum transport in protostellar disks \citep{Harsono10}, an effect
missing in the current two-dimensional disk model. The net result is some decrease in the 
disk mass and an equivalent increase in the stellar mass.

\section{Conclusions}
\label{summary}
Using numerical hydrodynamics simulations in the thin-disk approximation, we model the formation 
and evolution of protostellar 
disks in the embedded phase of star formation (EPSF) when the disks are exposed to intense mass 
loading from natal cloud cores. A wide spectrum of initial core masses and ratios $\beta$ of the 
rotational to gravitational energy is considered,
which allows us to construct statistical relations between disk masses $M_{\rm d}$, disk outer 
radii $r_{\rm d}$, and stellar masses $M_\ast$. Typical gas surface densities and midplane
temperatures at 40~AU and 100~AU are derived and analyzed as a function of stellar mass. 
Our findings can be summarized as follows.

\begin{itemize}

\item Our numerical modeling reveals that most of the Class 0 and I objects with stellar masses 
$M_\ast>0.05-01~M_\sun$ are characterized by disk masses
and gas surface densities greater than that of the minimum mass solar nebular, the latter defined
as in \citet{Hayashi85}. At the same time, disk radii for these objects start to grow beyond 100~AU,
making gravitational fragmentation possible. These results suggest
that giant planet formation  may commence as early as in the Class~0 phase, substantiating 
an earlier conclusion of \citet{Greaves10} based on the statistics of observationally inferred 
disk masses in different stages of star formation.

\item Disk masses in the EPSF show a near-linear correlation with stellar masses. The correlation 
may weaken somewhat if more models with low $\beta$ are considered but is unlikely to break completely.

\item A wide scatter of disk masses for a given stellar mass is likely caused by the 
distribution in the initial conditions 
of cloud cores and, in particular, by a wide spectrum of initial rotation rates.

\item Our mean and median disk masses in the Class~0 and I phases
($M_{\rm d,C0}^{\rm mean}=0.12~M_\odot$, $M_{\rm d,C0}^{\rm mdn}=0.09~M_\odot$ 
and $M_{\rm d,CI}^{\rm mean}=0.18~M_\odot$, $M_{\rm d,CI}^{\rm mdn}=0.15~M_\odot$, respectively)   
are {\it greater} than those inferred from observations by (at least)
a factor of 2--3. We show that this difference may be caused by high opacity in 
the inner disk regions.

\item Disk-to-star mass ratios are found to lie mostly in the $0.2<\xi<1.0$ range, with the mean
and median values $\xi\approx 0.65-0.7$ in the Class 0 and I phases.  Objects with 
$\xi<0.2$ are rare due to fast accumulation of disk mass in the early Class~0 phase, 
while objects with $\xi>1.0$ quickly evolve into binary or multiple systems, thus 
effectively reducing the disk mass. The least-squares fitting reveals little correlation between 
$\xi$ and $M_\ast$ in the EPSF.

\item Our modeling predicts a sub-linear (close to a square root) scaling between time-averaged disk
outer radii and stellar masses. Disks grow in size in the EPSF
due to mass loading from collapsing cores. Our mean and median values in the Class 0 and I phases 
($r_{\rm d,C0}^{\rm mean}=230$~AU, $r_{\rm d,C0}^{\rm mdn}=140$~AU   and $r_{\rm d,CI}^{\rm mean}=480$~AU,
$r_{\rm d,CI}^{\rm mdn}=290$~AU, respectively) 
suggest that  protostellar disks are large enough for giant planets to form via disk 
fragmentation.

\item We find sub-linear correlations between gas surface densities at 40~AU and 100~AU 
and stellar masses and also a square-root scaling between disk midplane temperatures and 
stellar masses, implying that massive disks are denser and hotter than their low-mass counterparts.
\end{itemize}

Although disk conditions seem to be favorable for giant planets to start forming as early as 
in the EPSF, we want to stress
that the direct gravitational instability scenario may encounter serious difficulties caused
by rapid inward migrations of the forming protoplanetary embryos \citep{VB06,VB10b,Nayakshin10,Cha10}.
According to recent numerical hydrodynamics simulations by \citet{VB10a}, the fraction of systems 
with survived embryos amounts to 10\% at best.

E.I.V. is thankful to the anonymous referee for useful suggestions that helped to 
improve the manuscript and gratefully acknowledges present support 
from an ACEnet Fellowship. Numerical simulations were done 
on the Atlantic Computational Excellence Network (ACEnet),
and at the Center of Collective Supercomputer
Resources, Taganrog Technological Institute at Southern Federal University.
This project was also supported by RFBR grant 10-02-00278 and by the 
Ministry of Education grant RNP 2.1.1/1937.

\end{document}